\documentstyle[preprint,aps]{revtex}
\begin{document}
\input{psfig.sty}%
\def \gam {\frac{ N_f N_cg^2_{\pi q\bar q}}{8\pi} }
\def \gamm {N_f N_cg^2_{\pi q\bar q}/(8\pi) }
\def \be {\begin{equation}}
\def \ba {\begin{eqnarray}}
\def \ee {\end{equation}}
\def \ea {\end{eqnarray}}
\def \gap {{\rm gap}}
\def \gaps {{\rm {gaps}}}
\def \gappp {{\rm \overline{\overline{gap}}}}
\def \im {{\rm Im}}
\def \re {{\rm Re}}
\def \Tr {{\rm Tr}}
\def \P {$0^{-+}$}
\def \S {$0^{++}$}
\def \uu {$u\bar u+d\bar d$}
\def \ss {$s\bar s$}

\title{ 
Comparing the Broken U3$\times$U3 Linear Sigma Model with Experiment.
}
\author{Nils A T\"ornqvist        \\
{\em Physics Dept. University of Helsinki, PB9, Fin-00014 Helsinki, Finland} \\
}
\maketitle
\begin{abstract}
The linear $\sigma$ model with broken U3$\times$U3 is compared with data
on the lightest scalar and pseudoscalar mesons.
When 5 of the 6 parameters  are fixed by the 
pseudoscalar masses and decay constants one  finds that,
already at the tree level, a reasonable description
for the 4 scalar masses, mixing 
and up to 8 tri-linear couplings of lightest scalars,
taken as $a_0(980),\ f_0(980),\ \sigma(\approx 500) $ and $K^*_0(1430)$.
This clearly indicates that these scalars are the chiral partners of the
$\pi ,\ \eta ,\ \eta ' ,\ K$ and strongly suggests 
that they like the latter are (unitarized) $q \bar q$ states.
\end{abstract}
\baselineskip=22pt

\section{Introduction}

As is well known the naive quark models (NQM) fails badly in 
trying to understand the lightest  scalars, the 
$a_0(980),\ f_0(980),\ K^*_0(1430)$ and the $\sigma(400-1200)$,
which we shall here call $\sigma(500)$.
Therefore today most authors want to give the 
$a_0(980),\ f_0(980)$ and the $\sigma(500) $ other interpretations 
than being $q\bar q$ states. Popular alternative interpretations are 
$K\bar K$ bound states, 4 quark states,
or for the $\sigma$, a glueball.
But there is also an obvious reason for  why the NQM fails:
Chiral symmetry is absent in the NQM, but is crucial for the scalars.
Chiral symmery is widely believed to be broken in the vacuum, and
 two of the scalars ($\sigma$ and $f_0$) have the same quantum 
numbers as the vacuum. 
Thus to understand the scalar nonet in the same way as we 
believe we understand the vectors and heavier multiplets,
and to make a sensible comparison with experiment, 
one must include chiral symmetry in 
addition to flavour symmetry into the quark model.
 
The simplest such chiral quark model is the 
linear U3$\times$U3 sigma model with 3 flavours.
Then we can treat both the scalar and pseudoscalar nonets simultaneously, and
on the same footing, getting automatically small masses for the pseudoscalar 
octet,
and symmetry breaking through the vacuum expectation values (VEV's) of the 
scalar fields.

As an extra bonus we have in principle a 
renormalizable theory, i.e. ``unitarity 
corrections'' are calculable.  
In fact, in the flavour symmetric ($u=d=s$ below) 
limit many of the unitarity corrections can 
be considered as being already included 
into the mass parameters of the theory, once the original 4-5 parameters are
replaced by the 4 physical 
masses for the singlet and octet $0^{-+}$ and $ 0^{++}$ masses,
and the pseudoscalar decay constant. 

Unfortunately this over 30 years old 
model\cite{sigma} has had very few phenomenological 
applications, although  important exceptions are the intensive efforts of 
M. Scadron and collaborators\cite{scad}. The 
reestablishment\cite{pdg98,NATfrasc} 
of the light and broad $\sigma$  has also more recently
revitalized the interest in the linear sigma 
model\cite{napsu,gavin,sigmafits}.  

\section{The Linear sigma model with 3 flavours}

The well known linear sigma model\cite{sigma} generalized to 3 flavours
 with complete  scalar ($s_a$) and pseudoscalar ($p_a$)
nonets has at the tree-level the 
Lagrangian  the same flavour and chiral symmetries 
as massless QCD. The  U3$\times$U3 Lagrangian with a symmetry breaking term 
${\cal L}_{SB}$ is  
\be 
 {\cal L}=
\frac 1 2 \Tr [\partial_\mu\Sigma \partial_\mu\Sigma^\dagger]
-\frac 1 2 \mu^2\Tr [\Sigma \Sigma^\dagger] -\lambda \Tr[\Sigma\Sigma^\dagger
\Sigma\Sigma^\dagger]\ -\lambda' 
(\Tr[\Sigma\Sigma^\dagger])^2+{\cal L}_{SB} \ .
\label{lag}
\ee

Here   $\Sigma$ is a $3\times 3$ complex
matrix, $\Sigma=S+iP= \sum_{a=0}^8(s_a+ip_a)\lambda_a/\sqrt 2$, in which
$\lambda_a$ are the Gell-Mann matrices, normalized as $\Tr[\lambda_a\lambda_b]=
2\delta_{ab}$, and where for the singlet 
$\lambda_0 = (2/N_f)^{1/2} {\bf 1}$ is 
included. Each  meson in Eq. (1) has a definite SU3$_f$ symmetry
content, which in the quark model means that it has the same  flavour
structure as
a  $q\bar q$ meson. Thus the fields $s_a$ and $p_a$ and 
potential terms in Eq.~(1) can be given
a conventional quark line structure\cite{NATPL} (Fig. 1). 

\begin{figure}[h]
\centerline{
\protect
\hbox{
\psfig{file=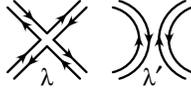,width= .15\textwidth,angle=0}}}
\caption{Graphical quark line 
representation of the $\lambda$ and $\lambda '$ terms
of Eq.~(1) }
\end{figure}

The symmetry breaking terms are most simply:
\be
{\cal L}_{SB}=\epsilon_\sigma \sigma_{u\bar u+d\bar d} + 
\epsilon_{s\bar s} \sigma_{s\bar s} +
\beta [\det \Sigma +\det \Sigma^\dagger]\ ,
\ee 
which  give the pseudoscalars mass and  break the 
flavour and $U_A(1)$ symmetries. 
The stability condition, that the linear terms
in the fields must vanish after the shift of the scalar fields 
($\Sigma\to \Sigma +V$) determines
the small parameters $\epsilon_i$ in terms of 
the pion and kaon masses and decay constants. One finds 
 $\epsilon_\sigma = m_\pi^2f_\pi$, $\epsilon_{s\bar s}=
(2m_K^2f_K-m_\pi^2f_\pi)/\sqrt 2$, 
while $\beta$ in the $U_A(1)$ breaking term is 
determined by $m_{\eta '}$, or by 
$m^2_\eta+m^2_{\eta '}$.

My fit to the scalars with the unitarized quark model (UQM)\cite{NAT} 
is essentially a  unitarization of 
eq.(\ref{lag}) with $\lambda\approx 16$ and $\lambda '= 0$, 
and with the main symmetry breaking generated by 
putting the pseudoscalar masses  at their physical values.
The  model was used as an effective theory 
with a symmetric smooth 3-momentum cutoff 0.54 GeV/c given by a gaussian 
form factor. Such a form factor is natural, since physical mesons are of 
course not pointlike, but have a finite size of 0.7-0.8 fm. 
(See  the discussion in connection to Eq.(\ref{width}) below.)
The fit  included the Adler zeroes which follow from eq.(1),
but only approximate crossing symmetry. 

Here I shall study the theory 
 at the tree level, leaving the detailed discussion of the
unitarization for future work.  In fact, when tadpole loops are 
included in the unitarization the ``unitarity corrections'' to the masses
should not be too large, since the tadpole loops
partly cancel the ($\log \Lambda$ divergence in) $s$-channel hadron loops.
One expects the corrections to the mass spectrum to be 
at most of the same order as the flavour symmetry breaking, 
since because of the renormalizability,
one can  in the flavour symmetric limit,
include the unitarity corrections into the mass parameters, fixed by
experiment. 

Eq.~(1) without ${\cal L}_{SB}$ 
is clearly invariant under $\Sigma \to U_L\Sigma U_R^\dagger$ of
U3$\times$U3.
After  shifting the flavourless scalar fields by  
the VEV's ($\Sigma \to \Sigma+ V$) to the minimum of the potential, 
the scalars aquire masses and also 
the pseudoscalars obtain a  (small) mass because of  
${\cal L}_{SB}$. Then the $\lambda$ and $\lambda'$ terms  
generate trilinear $spp$ and $sss$
couplings, in addition to those coming from the $U_A(1)$ symmetry breaking 
determinant term. The $\lambda$ term, which turns out to be the largest, 
obeys the OZI rule, 
while the $\lambda '$ and $\beta$ terms violate this rule. 
 
\section{ Tree-level masses.}

It is an ideal problem for a symbolic program like Maple V to calculate the
predicted masses, and couplings from the Lagrangian, which has 
6 parameters, $\mu,\lambda,\lambda',\beta$, $u=d$ and $s$, of which the
last two define the diagonal 
matrix $V$ with the flavourless meson VEV's: $V=diag[u,d,s]$. These are at the 
tree level related to the pion and kaon decay constants through
 $u=d=<\sigma_{u\bar u,d\bar d}>/\sqrt 2=f_\pi/\sqrt 2$ (assuming 
isospin exact) and $s=<\sigma_{s\bar s}>=(2f_K-f_\pi)/\sqrt 2$.
One finds  denoting the often occurring combination 
$\mu^2+4\lambda'(u^2+d^2+s^2)$ by $\bar \mu^2$, 
and expressing the flavourless mass matrices in the ideally mixed frame:
\ba
m^2_{\pi^+}\ &=&\bar \mu^2 + 4\lambda(u^2+d^2-ud)+2\beta s \\
m^2_{K^+}  \ &=&\bar \mu^2 + 4\lambda(u^2+s^2-su)+2\beta d \\
\begin{array}{c} m^2_\eta \\ m^2_{\eta'} \end{array} \ &=& eigv
\left( \begin{array}{cc} \bar \mu^2+2\lambda(u^2+d^2)-2\beta s & -\beta 
\sqrt 2 (u+d) \\
-\beta \sqrt 2 (u+d)& \bar \mu^2+4\lambda s^2 \end{array}\right) \\
m^2_{a_0^+}\   &=&\bar \mu^2 + 4\lambda(u^2+d^2+ud)-2\beta s \\
m^2_{\kappa^+} \  &=&\bar \mu^2 + 4\lambda(u^2+s^2+su)-2\beta d \\
\begin{array}{c} m^2_\sigma \\ m^2_{f_0} \end{array}\  &=& eigv
\left( \begin{array}{cc} \bar \mu^2+4\lambda'(u+d)^2+6\lambda(u^2+d^2)+2+
\beta s
 & (4\lambda' s+\beta)\sqrt 2 (u+d) \label{masses}\\
(4\lambda' s+\beta )\sqrt 2 (u+d)& \bar \mu^2+
8\lambda' s^2+12\lambda s^2 \end{array}\right) \\
\phi^{s\bar s-\eta '}&=&
\frac 1 2 \arctan\frac{-2\sqrt 2\beta (u+d)}{2\lambda(u^2+d^2-2s^2)-2\beta 
s} \\
\phi^{s\bar s-f_0}&=&\frac 1 2 \arctan\frac{2 \sqrt 2(4\lambda ' s+\beta 
)(u+d)}
{4\lambda '[(u+d)^2-2s^2]+6\lambda (u^2+d^2-2s^2)+2\beta s} 
\ea
where $eigv$ means the eigenvalues of the matrix which follows.

Let us first discuss the flavour symmetric limit $u=d=s$. Then  
 the pseudoscalar decay constants are equal $f_P=\sqrt 2 u= f_\pi=f_K$, while
the mixing angles $\Theta^{\eta'-singlet}=\phi^{s\bar s-\eta '}-54.73^\circ$
 and $\Theta^{\sigma-singlet}=\phi^{s\bar s-\sigma}+35.26^\circ$ 
vanish, and one has 4 nondegenerate physical masses for the octet and singlet
mesons $m_{P8},m_{S8},m_{P0},m_{S0}$.
Then there are also simple relations between the 5 model parameters 
$\lambda , \lambda ',
\bar \mu^2, \beta $ and $u=d=s$ and the 4 physical masses and the 
decay constant $f_P$:
\ba
\lambda    &=& (3m^2_{S8}-2m^2_{P0}-m^2_{P8})/(12f_P^2)         \\
\lambda '  &=& ( m^2_{S0}-m^2_{S8}-m^2_{P8}+m^2_{P0})/(12f_P^2) \\
\bar \mu^2 &=& (-3m^2_{S8}+5m^2_{P8}+4m^2_{P0})/6       \label{couplings} \\
\beta          &=&(m^2_{P8}-m^2_{P0})/(3\sqrt 2 f_P)                           \\
u=d=s &=& 
<\sigma_{u\bar u}>=<\sigma_{d\bar d}>=<\sigma_{s\bar s}>=f_P/\sqrt 2 
\ea
It is obvious that we can reparametrize the theory in terms of these physical 
quantities, which can be kept fixed in the renormalization, as long as 
flavour symmetry is exact. (If one choses the same tree-level 
values for the parameters
$\lambda , \mu^2 , \beta $ 
as found below in Eq.(\ref{param}) and $\lambda '=1$ but for 
$u=d=s$ the average value, 75.05 MeV, or $f_P=106$ MeV, one would have 
$m_{P8}= 384$ MeV, $m_{P0}=956$ MeV, 
$m_{S8}=1086$ MeV, $m_{S0}=741$ MeV.) 
Thus these masses can be thought of as already ``unitarized''.
On the other hand the original parameters and
induced tri-meson couplings  will be renormalized from the tree level values.

 Now breaking  the flavour symmetry ($s\ne u=d$)  we have only one more 
parameter, given by $f_K-f_\pi$, and it is evident that this breaking splits
the degeneracy in the mass spectrum from 4 independent masses 
to 8 masses and generates two mixing angles. Thus  one gets several tree-level
predictions in particular for the scalars (Table I).
 Of course now we expect these 
tree-level predictions to receive corrections from the unitarization,
but for  small symmetry breaking  (experimentally $(s-u)/s\approx 0.308$)
one would expect these corrections not to be larger than this, i.e. $<$30\%. 

 We can fix 5 of the 6 parameters, leaving  $\lambda'$ free,
 by the 5 experimental 
quantities from the pseudoscalar sector alone:
$m_\pi$, $m_K$, $m^2_\eta+m^2_{\eta'}$,
$f_\pi=92.42$ MeV and $f_K=113$ MeV\cite{pdg98}, which all are accurately known from experiment. One finds that at the tree level 
\be \lambda=11.57,\  
\bar \mu^2=0.1424\ {\rm GeV}^2,\ \beta =-1701\ {\rm MeV, }
\ u=d=65.35\ {\rm MeV},\  s=94.45\ {\rm MeV}.\label{param}\ee
The remaining $\lambda'$ parameter changes only the $\sigma$ and $f_0$ masses 
and their trilinear couplings, 
not those of the pseudoscalars. Gavin et. al.\cite{gavin}
calls for a group theoretical reason for this simplificati on.
In fact, graphically it is easy to see that the $\lambda '$ term 
can only break the OZI rule for the scalars, which can couple to the vaccuum, 
but for the pseudoscalars the $\lambda '$ term must leave the OZI rule intact.
Therefore  $\lambda '$ and $\mu$ affects the pseudoscalars only 
through the combination $\bar \mu^2=\mu^2+4\lambda '(u^2+d^2+s^2)$.

It is of some interest that the simple Gell-Mann--Okubo mass formula for the
mixing does not give the same result for the mixing angle between $\eta$ 
and $\eta '$ as our  model.
 This is because there is flavour symmetry breaking also in the
anomaly terms $\beta d$ and $\beta s$ in Eqs. (3-10) above. E.g. for the octet 
pseudoscalar mass one gets from Eqs. (3-10) $(4m_K^2-m_\pi^2)/3=542.5$ MeV
and a mixing angle of $\Theta^{s\bar s- \eta '}=-12.7^\circ$. 
This would be closer to the conventional mixing angle ($-10^\circ$ to 
$-23^\circ$)\cite{pdg98}, than our model 
$-5.0^\circ$, where the  octet $\eta_8$ mass is 566.1 MeV.    

Some of 
the couplings of $\sigma $ and $f_0$ depend sensitively on  $\lambda '$, 
since $\lambda '$ changes the small ideal mixing angle, $\phi^{s\bar s-f_0}$. 
It turns out below that $\lambda '$ must be small,
compared to $\lambda$, in order to fit the tri-linear couplings. 
By putting $\lambda'=1$ one gets a reasonable compromise for most of 
these couplings. With $\lambda'\approx 3.75$ one almost 
cancels the OZI rule breaking 
coming from the determinant  term, and the scalar mixing becomes near ideal
(for $\lambda '=-\beta /(4s)=4.5$ the cancellation is  exact).

As can be seen from Table I the predictions 
 are not far from the experimental masses taken as $a_0(980),\ f_0(980),\
K^*_0(1430)$, and $\sigma(500)$. For a discussion of the existence of the
$\sigma(500)$  see my recent Frascati 
talk\cite{NATfrasc}, which also includes some preliminary results of the present paper. 
Considering that one expects from our previous UQM 
analysis of the scalars\cite{NAT}
and our discussion above  that unitarity 
corrections can  be up to 30\% , and should 
go in the right direction compared to experiment, one must
conclude that these results are even better than expected. 
Similar mass analyses as in Table I have been done 
in Refs.\cite{gavin,sigmafits}, although with 
somewhat different input data.

\section{ Tri-linear couplings at the tree level.}

The trilinear coupling constants follow from the Lagrangian after 
one has made the shift $\Sigma\to\Sigma+V$. The most important
spp couplings  are at the tree level, when expressed in terms of the original
parameters:
\ba
g_{\kappa^+K^0\pi^+}&=&4\lambda (d-u+s)-2\beta \\
g_{\kappa^+ K^+\eta}\ &=&-4\lambda u\sin \phi^{s\bar s-\eta '}+
(2\sqrt 2 \lambda s+\beta \sqrt 2) \cos \phi^{s\bar s-\eta '} \\ 
g_{\kappa^+ K^+\eta '}\ &=&4\lambda u\cos \phi^{s\bar s-\eta '}+(2\sqrt 2\lambda s
+\beta \sqrt 2 )\sin \phi^{s\bar s-\eta '} \\
g_{\sigma\pi^+\pi^-}&=&2\sqrt 2 \cos \phi^{s\bar s-f_0} (u+d)[\lambda +
2\lambda '] -\sin \phi^{s\bar s-f_0}(8\lambda 's+2\beta ) \\
g_{\sigma K^+K^-}&=& \cos \phi^{s\bar s-f_0} \sqrt 2[\lambda (4u-2s)+
4\lambda ' (u+d)+\beta ]+ 4\sin \phi^{s\bar s-f_0}[\lambda (u-2s)-2\lambda ' s ] \\
g_{f_0\pi^+\pi^-}&=&2\sqrt 2\sin \phi^{s\bar s-f_0} (u+d)[\lambda +
2\lambda '] +\cos \phi^{s\bar s-f_0}(8\lambda 's+2\beta ) \\
g_{f_0    K^+K^-}&=& \sin \phi^{s\bar s-f_0} \sqrt 2[\lambda (4u-2s) +
4\lambda ' (u+d)+\beta ]- 4\cos \phi^{s\bar s-f_0}[\lambda (u-2s)-2\lambda ' s ] \\
g_{a_0\pi\eta}&=&\cos \phi^{s\bar s-\eta '}2\sqrt 2\lambda (u+d) -
2\beta \sin\phi^{s\bar s-\eta '}\\
g_{a_0\pi\eta'}&=&\sin \phi^{s\bar s-\eta '}2\sqrt 2\lambda (u+d) +
2\beta \cos\phi^{s\bar s-\eta '}\\
g_{a_0 K^+K^-}&=&\sqrt 2[\lambda (4u-2s)-4\lambda '(d-u)-\beta ] 
\ea
In fact these can be written in  more useful forms 
in terms of the predicted physical masses and mixing
angles and decay constants:
\ba
g_{\kappa^+K^0\pi^+}&=&(m^2_\kappa-m^2_\pi)/(\sqrt 2 f_K) \\
g_{\kappa^+ K^+\eta}\ &=&-\sqrt 3\sin( \phi^{s\bar s-\eta '}-35.26^\circ )
(m^2_{\kappa}-m^2_\eta)/(2f_K) \\
g_{\kappa^+ K^+\eta'}\ &=&\sqrt 3\cos( \phi^{s\bar s-\eta '}-35.26^\circ )
(m^2_{\kappa}-m^2_{\eta'})/(2f_K) \\  
g_{\sigma\pi^+\pi^-}&=&\cos \phi^{s\bar s-f_0}(m^2_\sigma-m^2_\pi)/f_\pi \\
g_{\sigma K^+K^-}&=&-\sqrt 3\sin( \phi^{s\bar s-f_0}-35.26^\circ )
(m^2_{\sigma}-m^2_K)/(2f_K) \\
g_{f_0\pi^+\pi^-}&=&\sin \phi^{s\bar s-f_0}(m^2_{f_0}-m^2_\pi)/f_\pi
 \label{f0pipi}\\
g_{f_0 K^+K^-}&=&\sqrt 3\cos( \phi^{s\bar s-f_0}-35.26^\circ )
(m^2_{f_0}-m^2_K)/(2f_K) \\
g_{a_0\pi\eta}&=&\cos \phi^{s\bar s-\eta '}(m^2_{a_0}-m^2_\eta)/f_\pi \\
g_{a_0\pi\eta'}&=&\sin \phi^{s\bar s-\eta '}(m^2_{a_0}-m^2_{\eta'})/f_\pi \\  
g_{a_0 K^+K^-}&=&(m^2_{a_0}-m^2_K)/f_K 
\ea

In table II 8 different spp couplings are compared with quoted 
experimental numbers. 
In some of the channels of  table II the resonance is 
below threshold and the widths therefore vanish at the resonance mass. 
However, the coupling
constants have recently been determined through a loop diagram from
$\phi\to K\bar K\to\gamma\pi\pi$ and $\phi\to K\bar K\to\gamma\pi\eta$ (albeit in a somewhat model 
dependent way) by the Novosibirsk group\cite{ach,nov}.
For channels where the phase space is large, it is important that one includes
a form factor related to the finite size of physical mesons. In the 
$^3P_0$ quark pair creation  model a radius of $\approx$0.8 fm 
leads to a gaussian form factor,
as in the formula below, where $k_0 \approx 0.56$ GeV/c (as was found
 in the UQM\cite{NAT}). 
Thus the  widths are computed from the formula: 
\be
\Gamma(m)=\sum_{isospin}\frac{g_i^2}{8\pi}
\frac{k^{cm}(m)}{m^2}e^{-[k^{cm}(m)/k_0]^2}\label{width}\ .
\ee

As can be seen from table II most of the  couplings 
are not far from experiment.
Only the $f_0\to\pi\pi$ and $a_0\to\pi\eta$ couplings and widths
come out a bit large, but these are very sensitive to higher order loop  
corrections due to the $K\bar K$ threshold, and $f_0\to\pi\pi$ is 
extremely sensitive to the scalar near-ideal mixing angle and $\lambda '$.
 If one choses $\lambda'= 3.75$  this mixing angle nearly vanishes 
($\phi^{s\bar s-f_0}=-3.0^\circ$) together with the $f_0\to\pi\pi$ coupling 
(c.f. eq.(\ref{f0pipi})).
From our experience with the UQM\cite{NAT} the $a_0\to K\bar K$ peak width, 
when unitarized, is reduced, because of
 the $K\bar K$ theshold, by up to a factor 5. Therefore one cannot 
expect that the tree level couplings should agree better with data 
than what those of Table II do.  After all, this is a very strong
coupling model ($\lambda=11.57$, leading to large $g_i^2/4\pi$)
and higher order effects should be important.

\section{Conclusions.}
In summary, I find that the linear sigma model 
with three flavours, at the tree level, works much  
better than what is generally believed.
 When the 6 model parameters are fixed mainly by the
pseudoscalar masses and decay constants,
 one predicts the 4 scalar masses
and mixing angle to be near those of the experimentally observed nonet $
a_0(980),$ $ f_0(980),$ $\sigma(500),$ $ K^*_0(1430)$.
Also 8 couplings/widths of the scalars to two pseudoscalars are predicted
reasonably close to  their  presently known, 
rather uncertain experimental values. The agreement is good enough considering
that 
some of these are expected to have large higher order corrections. The model
 works, in my opinion, just as well as the 
naive quark model works for the heavier nonets.
A more detailed data comparison would become meaningful,
after one  has included  higher order effects, i.e. after one 
has  unitarized the model, 
e.g., along the lines of the UQM\cite{NAT}. 

Those working on chiral perturbation theory and nonlinear sigma models
 usually point out that
 the linear model does not predict all low energy constants 
correctly. However, one should remember that the energy regions of validity
are different for the two approaches.
 Chiral perturbation theory usually breaks down when one approaches
the first scalar resonance. 
The  linear sigma model, on the other hand,
includes the scalars from the start 
and can be a reasonable interpolating model in the intermediate 
energy region near 1 GeV, where QCD is too difficult to solve.

These results strongly favour the interpretation that the 
$a_0(980)$, $f_0(980)$, $\sigma(500)$,  $ K^*_0(1430)$  
belong to the same nonet, 
and that they are the chiral partners of the $\pi$, $\eta$,  $\eta '$, $ K$. 
If the latter are believed to be unitarized $q\bar q$ states, 
so are the light scalars  $
a_0(980),$ $ f_0(980),$ $\sigma(500),$ $ K^*_0(1430)$, and the broad 
$\sigma(500)$ should be interpreted as an existing resonance.
The $\sigma$ is a very important hadron indeed, 
as is evident in the sigma model, because this is the boson
which gives the constituent quarks most of their mass
and thereby it gives also 
the light hadrons most of  their mass. Therefore it is natural to consider
the $\sigma (500)$ as the Higgs boson of strong interactions. 

{\it Acknowledgement.} This work is partially supported by
 the EEC-TMR program Contract N.CT98-0169.

\begin{table}[t]
\centering
\caption{ \it Predicted masses in MeV and mixing angles for two values of the 
 $\lambda'$ parameter. The 
 asterix means that $m_\pi,m_K$
and $m_\eta^2+m^2_{\eta'}$ are fixed by experiment together with $f_\pi=$92.42 MeV and 
$f_K=$113 MeV.  }
\vskip 0.1 in
\begin{tabular}{|l|c|c|c|} \hline
Quantity       &  Model $\lambda '=1$& Model $\lambda '=3.75$& Experiment \\
\hline
\hline
$m_\pi$   &  137$^{*)}   $& 137$^{*)}   $ &137\cite{pdg98}  \\
$m_K$     &  495$^{*)}   $& 495$^{*)}   $ &495\cite{pdg98}  \\
$m_\eta  $&  538$^{*)}   $& 538$^{*)}   $ &547.3\cite{pdg98}  \\
$m_{\eta'}$ &963$^{*)}   $& 963$^{*)}   $ &957.8\cite{pdg98}  \\ 
$\Theta^{\eta'-singlet}$ &-5.0$^\circ$ &-5.0$^\circ$ &(-16.5$\pm6.5)^\circ$\cite{pdg98} \\ 
$m_{a_0}$ &1028 &1028 &  983\cite{pdg98}  \\ 
$m_{\kappa}$ &1123 &1123&   1430\cite{pdg98}  \\ 
$m_{\sigma}$ &651 &619  & 400-1200\cite{pdg98}  \\ 
$m_{f_0}$ &1229 &1188   &980\cite{pdg98}  \\ 
$\Theta^{\sigma-singlet}$ 
& 21.9$^\circ$ & 32.3$^\circ$ &(28-i8.5)$^\circ$\cite{NAT}  \\
\hline
\end{tabular}
\label{tab1}
\end{table}

\begin{table}[t]
\centering
\caption{ \it Predicted couplings $\sum_i\frac{g_i^2}{4\pi}$ (in GeV$^2$)
, when $\lambda'=1$, 
compared with experiment and predicted widths with experiment (in MeV). (We 
have used isospin invariance to get the sum over charge channels, when there
is data for one channel only.)   The 
predicted $f_0\to \pi\pi$ width is extremely
 sensitive to the value of $\lambda^\prime$ (for $\lambda '= 
3.75 $  it nearly vanishes)
and unitarity effects as discussed in the text. Also the  
$a_0\pi\eta$ coupling 
is very sensitive to loop corrections due to the $K\bar K$ threshold.}
\vskip 0.1 in
\begin{tabular}{|l|c|c|c|c|} \hline
Process       &  $ \sum_i\frac{g_i^2}{4\pi}$ &  $ \sum_i\frac{g_i^2}{4\pi}$  &
$\sum_i\Gamma_i$ & $\sum_i\Gamma_i$ \\
      & in model &in experiment & model&experiment\\ 
\hline
\hline
$\kappa^+\to K^0\pi^++K^+\pi^0$   &  7.22 & -  & 678 & 
$278\pm23$\cite{pdg98,aston}   \\
$\kappa^+\to K^+\eta $     &  0.28 & $\approx 0\cite{aston}$& 13 & $<26$\cite{aston} \\
$\sigma\to \pi^+\pi^-+\pi^0\pi^0$   &  2.17 & 1.95\cite{ach} & 574 
&300-1000\cite{pdg98} \\
$\sigma\to K^+K^-+K^0\bar K^0$   &  0.16 & 0.004\cite{ach}   & 0   & 0\\
$f_0\to    \pi^+\pi^-+\pi^0\pi^0$   
&  1.67 & 0.765$^{+0.20}_{-0.14}$\cite{nov} & see text & 40 - 100\cite{pdg98}\\
$f_0\to    K^+K^-+K^0\bar K^0$   &  6.54 & 4.26$^{1.78}_{-1.12}$\cite{nov}   & 0   & 0\\
$a_0^+\to \pi^+\eta$   &  2.29 & 0.57\cite{nov}  & 273 see text & 50 - 100\cite{pdg98}\\
$a_0^+\to K^+\bar K^0$ &  2.05 &  1.34$^{+0.36}_{-0.28}$\cite{nov} &  0 &0  \\
\hline
\end{tabular}
\label{tab2}
\end{table}

\end{document}